\documentclass[default]{aastex701}
\usepackage{amsmath,epsfig}
\newcommand{\beq}{\begin{equation}}
\newcommand{\eeq}{\end{equation}}
\newcommand{\bea}{\begin{eqnarray}}
\newcommand{\eea}{\end{eqnarray}}
\newcommand{\rem}[1]{ }

\begin{document}
\title{Electron-positron cascade in magnetospheres of supermassive Kerr black holes and the origin of relativistic AGN jets}

\author[0000-0001-5987-2856]{Mikhail V. Medvedev} 
\affiliation{Department of Physics and Astronomy, University of Kansas, Lawrence, KS 66045}
\affiliation{Laboratory for Nuclear Science, Massachusetts Institute of Technology, Cambridge, MA 02139}
\email[show]{medvedev@ku.edu}
\author[0000-0002-0333-2452]{Andrzej A. Zdziarski} 
\affiliation{Nicolaus Copernicus Astronomical Center, Polish Academy of Sciences, Bartycka 18, PL-00-716 Warszawa, Poland}
\email[show]{aaz@camk.edu.pl}

\begin{abstract}
The avalanche mechanism of plasma production in active galactic nuclei (AGNs) is detailed, and constraints on system parameters required for efficient electron-positron pair cascades are explored. Whether an AGN falls within this favorable parameter range may explain the observed radio-loud versus radio-quiet dichotomy. However, this study shows that cascades produce orders of magnitude fewer pairs than are needed to explain the synchrotron emission observed in luminous jets. This suggests the existence of either an alternative lepton source, namely pair production of photons from the hot accretion flows around AGN central black holes, or matter loading of the jets from the surrounding medium, or, most likely, both. The case of the radio galaxy 3C 120 is considered in detail.
\end{abstract}
 
\section{Introduction}
\label{s:intro}

Collimated relativistic jets are observed in a wide variety of astrophysical objects that accrete plasma in the presence of magnetic fields. Jets originating from systems containing black holes (BHs) are likely associated with accretion flows. These jet-producing BH systems span a wide range of parameters. For instance, BH masses span nearly ten orders of magnitude, from stellar-mass BHs in binaries and $\gamma$-ray bursts to supermassive BHs in radio-loud and radio-quiet galaxies, quasars, and active galactic nuclei (AGNs). 

It is generally accepted that the physical mechanism responsible for jet production is the collimation of the ejected plasma by magnetic fields. To date, the mechanism identified for launching the most powerful jets is the Blandford-Znajek (BZ) mechanism, in which the rotational energy of a Kerr BH is tapped by large-scale magnetic fields threading the horizon \citep{BZ77}. This theoretical idea of extracting the BH's rotational energy electromagnetically has been confirmed by numerical simulations (see, e.g., \citealt{Semenov04, Tchekhovskoy11, Tchekhovskoy12}) and, possibly, by observations of stellar-mass BHs, through a claimed correlation between the radio luminosity of transient ballistic jets and BH spin parameters estimated via continuum fitting \citep{Narayan12, Steiner13}. 

The BZ mechanism relies on the formation of an e$^\pm$ cascade in the BH magnetosphere. In this cascade, seed electrons are accelerated by electric fields and then upscatter soft photons emitted by the surrounding accretion flow to energies above the pair-production threshold. Particle-in-cell simulations of such pair cascades in Kerr BH environments have been performed \citep{Parfrey19, Crinquand20, Yuan25}. These simulations demonstrate the feasibility of the pair-cascade model for explaining jet sources. In fact, the formation of such a cascade is a necessary condition for BZ jet formation. Without it, the vacuum magnetosphere yields null jet power \citep{Wald74}.  

On the other hand, the rate of e$^\pm$ pairs produced in the cascade is very low. It is usually parametrized by the ratio of the charge density in the magnetosphere to the Goldreich-Julian (GJ) density, whose order-of-magnitude estimate is 
\begin{equation}
    \rho_{\rm GJ}\sim \frac{(\frac{1}{2}\omega_{\rm BH}) B}{2\pi c},
    \label{nGJ}
\end{equation}
where $B$ is the magnetic field strength, $c$ is the speed of light, and $\omega_{\rm BH}$ is the angular velocity of BH rotation [see Eqs. \eqref{wBH}, \eqref{GJ} below for the exact definitions of $\omega_{\rm BH}$ and $\rho_{\rm GJ}$]. The density ratio, $\lambda\equiv \rho_{\rm e}/\rho_{\rm GJ}$, is called the pair multiplicity, and for magnetospheric cascades it is limited to $\lambda\lesssim \! 10^3$ \citep{Levinson11, Nokhrina15}. The corresponding lepton flux can be compared with that inferred from radio observations of luminous AGN jets \citep{Nokhrina15}. Those authors\footnote{\citet{Nokhrina15} reported $\lambda\gtrsim10^{12}$, but assumed that only 1 in 100 leptons undergo acceleration.} found large values, $\lambda\gtrsim10^{10}$. Magnetospheric cascades alone can account for the observed properties of only very weak sources, such as M87 and Sgr A$^*$ (e.g., \citealt{Levinson11, Moscibrodzka11}). This shows the necessity of another mechanism for lepton loading, either in the form of normal plasma or e$^\pm$ pairs\footnote{We note that this fact is often overlooked, with statements in the literature that BZ jets are by default pair-dominated, and a detection of baryons indicates a different type of jet (e.g., \citealt{Diaz_Trigo13}).}. 

A process likely to occur is e$^\pm$ pair production at the base of a Poynting-flux-dominated jet via collisions of energetic photons emitted by a hot accretion flow \citep{Henri91, B99_pairs, Levinson11, Moscibrodzka11, Aharonian17, Sikora20, Zdziarski22c}. Depending on the hot flow's power and spectral characteristics, this process can be substantially more efficient than the magnetospheric cascade. However, this does not rule out baryon loading from the jet's surroundings, e.g., \citet{ORiordan18}. In the case of the radio galaxy 3C 120, \citet{Zdziarski22c} calculated both the rate of pair production within the jet base (based on the hard-X-ray spectrum) and the rate of flow of the synchrotron-emitting leptons (based on the synchrotron spectrum of the core jet) and found them to be compatible with being the same. Similar calculations, showing that these two rates are comparable, were performed for the microquasars MAXI J1820+070 and Cyg X-1 \citep{Zdziarski22a, Z_Egron22}. Thus, at least in those sources, there is no need for substantial baryon loading to account for the synchrotron emission. On the other hand, some baryon loading is required, at least in some cases, to account for the jet energetics, e.g., \citet{Pjanka17}.

In this work, we re-examine the magnetospheric cascade process. We provide simple analytical estimates of its properties and derive the conditions under which it operates. We compare our results with previous numerical calculations. For 3C 120, we confirm that the rate of this process is many orders of magnitude lower than that inferred from the synchrotron spectrum. We also consider whether the well-known dichotomy between radio-loud and radio-quiet AGNs (e.g., \citealt{Sikora07}) can be related to the derived condition for cascade formation.

\section{Electron-positron cascade in Kerr BH magnetospheres}
\label{s:cascade}

An energetic, often relativistic, plasma is present in extreme astrophysical environments, including spinning neutron stars, magnetars, and spinning BHs of various masses, which drive powerful relativistic jets. Unlike a neutron star, where charges can be pulled from the surface and populate its environment, a BH horizon cannot be a source of plasma. The plasma source must therefore be within the magnetosphere, namely the external magnetic field (since a BH cannot have its own $B$-field), with self-generated plasma flowing away or falling in through the outer and inner light cylinders. 

In the $e^{\pm}$ avalanche model of pair production discussed here, the region where plasma is generated is called the ``{\it spark gap}'' (or just the ``{\it gap}''). Whether plasma can be produced in the magnetosphere determines whether a relativistic jetted outflow is present in the system. Astronomical observations show that both possibilities exist in nature. 

We should note that vacuum breakdown cannot occur, especially around supermassive BHs, because the induced electric fields are too weak: $E\lesssim B\sim 10^5\textrm{ G}\ll E_{\rm Schwinger}$, where
\begin{equation}
 E_{\rm Schwinger}\equiv \frac{m_{\rm e}^2 c^3}{e \hbar}\approx 4.4\times 10^{13}\,{\rm G}
\end{equation}
is the critical field strength for spontaneous vacuum pair production, $e$ and $m_{\rm e}$ are the electron charge and mass, respectively, and $\hbar=h/2\pi$ is the Planck constant. Yet the leptonic pair plasma avalanche can be produced via a cascade initiated by seed charges accelerated in the spark-gap region, which scatter off photons emitted by an accretion flow. The cascade is a threshold process that depends on several system parameters.

\begin{figure}[b!]
\center
\includegraphics[angle = 0, width = 0.45\columnwidth]{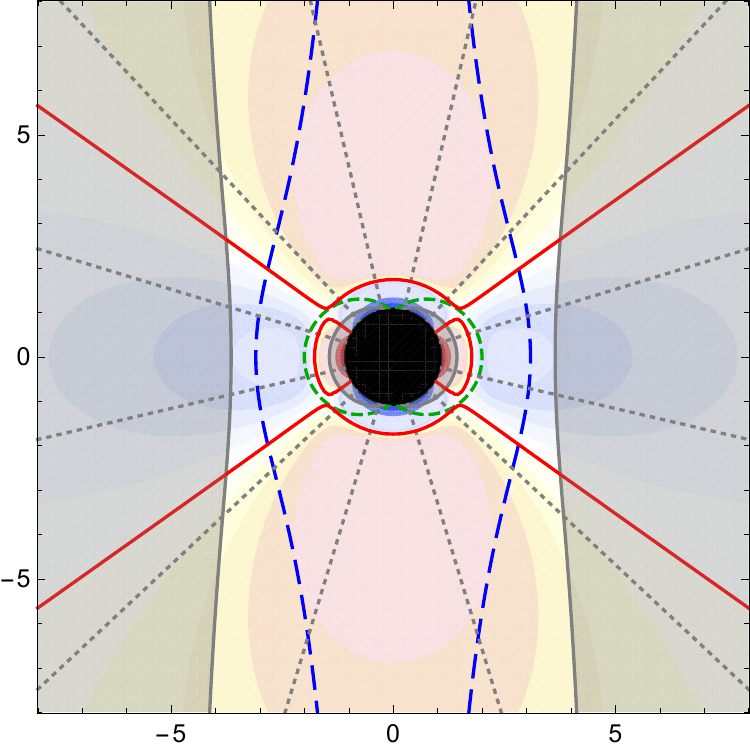}
\includegraphics[angle = 0, width = 0.441\columnwidth]{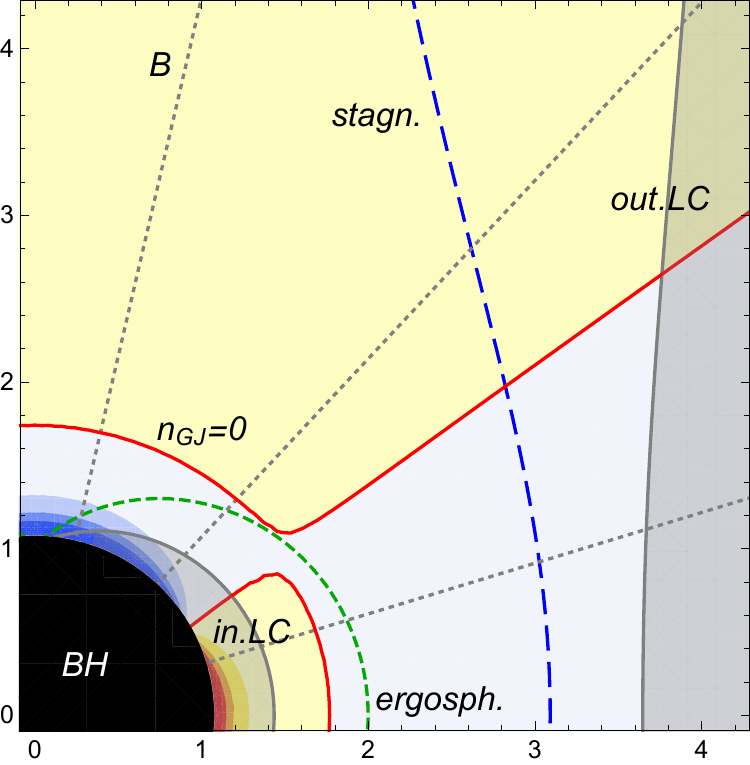}
\caption{Schematic representation of the accreting BH system. Shown are the ergosphere (green dashed curves), the inner and outer light cylinders (gray lines and shading), magnetic field lines (dotted lines), the stagnation surface (blue dashed curves), and the GJ null surface (red curves). Yellow-orange and blue indicate positive and negative charge densities, respectively. An accretion disk (not shown) lies in the equatorial plane. We assume a rapidly spinning BH with $a_*=0.99$ and $\omega_{\rm F}=0.45\omega_{\rm BH}$. Axis scales are in units of $r_{\rm g}$.}
\label{kerr-bh}
\end{figure}

To proceed, we need to set up the magnetic field geometry. If the plasma is sufficiently dense, the magnetosphere adopts a force-free configuration. The magnetic field takes a split-monopole solution (Figure \ref{kerr-bh}) of the Grad-Shafranov equation, with the current sheet in the equatorial plane \citep{Lyutikov12, Nathanail14}. The force-free condition breaks down in regions of weak (vanishing) magnetic field, e.g., in the current sheet. However, this is not considered a major problem because the solution, which becomes formally discontinuous, can be readily regularized by introducing finite plasma inertia, finite particle Larmor radius, collisionality, etc. 

For the force-free condition to hold, the plasma density must be sufficient to short-circuit the electric field component along the magnetic field (particle drifts of charges perpendicular to the $B$-field are slow and can be neglected). BHs have inner and outer light cylinders --- the imaginary surfaces through which plasma can move out of the magnetosphere inward or outward, respectively, as shown in Figure \ref{kerr-bh}.\footnote{Although the stagnation surface can be formally defined as the location where the inward gravitational pull is balanced by the radial component of centrifugal acceleration, it is of little importance. The plasma produced near the null surface leaves the region of its production with a finite hydrodynamic velocity.} These plasma flows should evacuate the entire region between them unless the plasma is replenished. If the plasma is not replenished in the required quantities, the magnetosphere is {\em not} force-free. Instead, it is represented by the vacuum solution \citep{Wald74}, which yields vanishing BZ jet power. In other words, the jet is nonexistent in a vacuum field configuration when the plasma replenishment process ceases.

\subsection{The cascade model}

Let us first assume that there is initially no plasma in a small region, called the spark gap region, of the BH magnetosphere near the null surface, where the GJ charge density vanishes, $\rho_{\rm GJ}=0$. Because this region is small, the absence of plasma does not violate the global force-free structure of the magnetosphere, even though the force-free approximation is generally invalid locally. Furthermore, if the gap size is small compared to its radial distance, $H\ll R$, the deviation from the force-free $B$-field geometry within the gap is negligible as well. Since the plasma is absent in the gap region, an uncompensated electric field can exist along the magnetospheric magnetic field. This electric field can accelerate charges to high energies, which can subsequently be converted into electron-positron pairs via the avalanche mechanism. Thus, the gap region is the natural site for abundant plasma production. 

The avalanche mechanism proceeds as follows. (i) A non-compensated parallel electric field accelerates a seed charge in the spark gap region. Provided the electric field and the gap size are sufficiently large, the charge is accelerated to relativistic energy, so its Lorentz factor is $\Gamma\gg1$. (ii) The charge propagates through the magnetosphere, a ``bath'' of soft photons copiously produced by the accretion disk. Incidentally, the charge can Compton-upscatter a soft background photon into a hard $\gamma$-ray photon, $e+\gamma_{\rm b}\to e+\gamma_{\gamma}$. (iii) Finally, a $\gamma$-ray photon interacting with a soft photon can produce an electron-positron pair, $\gamma_{\gamma}+\gamma_{\rm b}\to e^{+}+e^{-}$, provided that their total center-of-mass energy exceeds the electron rest energy, $m_{\rm e}c^{2}$. These secondary charges are accelerated by the parallel electric field, producing more $\gamma$-rays and more charges exponentially.

\subsubsection{Electric field and Lorentz factor}

Assuming a stationary, axisymmetric, degenerate force-free magnetosphere around a rotating BH, the electric field is purely poloidal \citep{Hirotani98}:
\begin{equation}
E_{\rm p}=-\frac{\omega_{\rm F}-\omega}{2\pi\alpha c}\nabla\Psi,
\end{equation}
where $\Psi$ is the magnetic flux, $\omega_{\rm F}$ is the angular velocity of the magnetic field lines, $\alpha$ is the redshift factor (also called the lapse function), and $\omega$ is the angular velocity of the zero angular momentum observer (ZAMO), who keeps their $r$ and $\theta$ coordinates constant. These parameters are defined in Kerr geometry, conventionally expressed in Boyer–Lindquist coordinates as 
\begin{align}
        & \alpha=\rho\Delta^{1/2}/\Sigma, \\
        & \omega=2 a r_{\rm g} r c/\Sigma^2,
\end{align}
where $r_{\rm g}=GM/c^2$ is the radius of the extremely spinning Kerr black hole, $\rho^2=r^2+a^2\cos^2(\theta)$, $\Delta=r^2-2r_{\rm g} r+a^2$, $\Sigma^2 =(r^2+a^2)^2-a^2\Delta\sin^2(\theta)$, and $a=J/Mc$ is the spin parameter \citep{Thorne86,Beskin92, Hirotani98}. Often, the dimensionless spin parameter $a_*= Jc/(GM^2) = a/r_{\rm g}$ is used, with $0 \le a_* \le 1$. Furthermore, the radius of an arbitrarily spinning BH is defined as the outer-horizon radius $r_{\rm BH}(a_*)=r_{\rm g}(1+\sqrt{1-a_*^2})$. We also note that the ZAMO angular velocity, $\omega$, vanishes at infinity ($\alpha\to1$) and coincides with the BH's uniform rotation \citep{MTW73, BZ77}. 
\begin{equation}
    \omega_{\rm BH}\equiv \frac{a_*c}{2r_{\rm BH}}
    =\frac{c}{2r_{\rm g}}\left(\frac{a_*}{1+\sqrt{1-a_*^2}}\right)
    \label{wBH}
\end{equation}
as one approaches the horizon, $\alpha\to0$.
The local GJ charge density is: 
\begin{equation} 
\rho_{\rm GJ}=\frac{1}{4\pi}\nabla\cdot E_{\rm p}=-\frac{1}{4\pi}\nabla\cdot \left(\frac{\omega_{\rm F}-\omega}{2\pi\alpha c}\nabla\Psi\right)~.
\label{GJ}
\end{equation}
Fig. \ref{kerr-bh} shows the distribution of the GJ density (yellow-red shades indicate $\rho_{\rm GJ}>0$ and blue shades indicate $\rho_{\rm GJ}<0$). It is clear that $\rho_{\rm GJ}$ changes sign, and the ``null surface" (where $\rho_{\rm GJ}=e n_{\rm GJ}=0$) is shown in red. Near the pole ($\theta\sim0$), this surface is nearly spherical and hence orthogonal to the magnetic field lines, since the toroidal component of the field is small there. This greatly simplifies the geometry, so we consider the polar region in our future estimates. The null surface is charge-deficient, leading to an electric field along the magnetic field lines, $E_{\|}$, which breaks the force-free condition. This results in the formation of a gap region (similar to the outer gap around a pulsar). Note that $\omega_{\rm F}$ is not generally known from first principles. For analytical estimates, one often chooses $\omega_{\rm F}=0.5\omega_{\rm BH}$, which maximizes the BZ power since it scales as $\omega_{\rm F}(\omega_{\rm BH}-\omega_{\rm F})$ (e.g., \citealp{KB14}). Numerical simulations indicate that the jet power, and hence $\omega_{\rm F}$, varies across different accretion systems (e.g., \citealp{Lowell2024}).  

Poisson's equation inside the gap is:
\begin{equation}
{dE_\|/dx}=4\pi[e(n^+-n^-)-\rho_{\rm GJ}] ,
\label{1}
\end{equation}
where $n^\pm$ is the number density of positrons and electrons, respectively, generated and moving in the gap. Assuming the gap is thin, one uses a 1D representation of the gap, with $x=0$ at the midpoint of the gap, where $\rho_{\rm GJ}=0$, and with $\hat{x}$ perpendicular to the null surface. Thus, $r_{0}$ is the radial position where $\rho_{\rm GJ}=0$, and the $x$-coordinate is $x=r-r_{0}$. Since the split-monopole field geometry is assumed, $x$ is also the coordinate along a field line (along which the charges move). 

The total width of the gap, $H$, where the parallel electric field is non-vanishing, is assumed to be small compared to the size of the extremely spinning Kerr BH, $r_{\rm g}\approx 1.5\times10^{13}\textrm{~cm}\, M/10^8M_\odot$, so one can expand $\rho_{\rm GJ}(x)$ around $x=0$. This expansion yields $\rho_{\rm GJ}(x)\simeq Ax$, where $A=\partial_x(\rho_{\rm GJ})$ evaluated at $x=0$. In general, the parameter $A$ is a complicated function of the BH mass and spin parameter, $M$ and $0\le a\le r_{\rm g}$, and of the azimuthal angle $\theta$ for a given distribution of the magnetospheric magnetic field near the horizon. From Poisson's equation, Equation \eqref{1}, we have $E_{\|}=4\pi A(x^{2}/2-H^{2}/8)$ inside the gap, $0\le |x| \le H/2$, and $E_{\|}=0$ at $x=\pm H/2$ and beyond. 

The electric field in the gap accelerates the charges. The motion of the charges as they are accelerated along the magnetic field lines can be approximated as one-dimensional \citep{Hirotani98}:
\begin{equation}
m_ec^2{d\Gamma}/dx=eE_\|-\frac{4}{3}(\Gamma^2-1)\sigma_{\rm T}U_{\rm b} ,
\label{2}
\end{equation}
where $\sigma_{\rm T}$ is the Thomson cross section, $\Gamma$ is the electron Lorentz factor, and $U_{\rm b}$ is the energy density of the background photons. The first term on the right-hand side is the acceleration due to the gap-parallel electric field, and the second is the Compton drag from the background photons. Note that this is valid only in the Thomson regime, not in the Klein-Nishina regime. This regime is generally valid for a broadband spectrum typical of accreting sources \citep{Beskin92}. In this context, we disregard curvature radiation, as it is negligible given the substantial curvature radius, which is comparable to or larger than the size of supermassive BHs (see e.g., \citealp{Beskin92}). 

\subsubsection{Inverse Compton scattering and pair production}

Under the assumptions that all motion is along the magnetic field lines and that the $e^\pm$ spectrum is monoenergetic at a given position, the continuity equations for $e^\pm$ are
\begin{equation}
\pm {d_x}\left[n^\pm(x)\beta(x)\right]=\int_0^\infty\eta_{\rm p}(\nu_\gamma)[F^+(x,\nu_\gamma)+F^-(x,\nu_\gamma)]d\nu_\gamma ~,
\label{3}
\end{equation}
where $\beta(x)c$ is the electron and positron velocity, and $F^\pm$ are the number densities of $\gamma$-rays propagating in the $\pm x$-directions. The pair-production redistribution function, $\eta_{\rm p}$, is the spectrum- and angle-averaged cross-section; it is discussed in detail in \citet{Hirotani98} and \citet{BLP82}. Using Equation\ \eqref{3}, one can see that the current in the gap along a field line is conserved along $x$, yielding an expression for the current density,
$e[n^+(x)+n^-(x)]c\beta(x)=j_0$.

The $\gamma$-ray fluxes $F^\pm(x,\nu_\gamma)$ for each $\gamma$-ray photon energy $h\nu_\gamma$ are given by the continuity equation:
\begin{equation}
\pm {\partial_x}F^\pm(x,\nu_\gamma)=\eta_{\rm C} n^\pm \beta(x)-\eta_{\rm p}F^\pm(x,\nu_\gamma) ~,
\label{4}
\end{equation}
where $\eta_{\rm C}$ is the Compton redistribution function (i.e., the spectrum-averaged cross-section), discussed in detail in \citet{Beskin92} and \citet{Hirotani98}. To fully describe the behavior of the charges and $\gamma$-rays in the gap, we need to specify a spectrum for the background photons. 

The system of equations \eqref{1}, \eqref{2}, \eqref{3}, \eqref{4}, supplied with appropriate boundary conditions, can be solved numerically. The results are reported in detail elsewhere \citep{Ford18, Sitarz24}. Their relevant aspects will be discussed where appropriate.

\subsection{Estimates}

Now, we present estimates that help us better understand the pair cascade and draw important conclusions. 

\subsubsection{Electric field}

We can write  the field strength as follows 
\begin{equation}
E_{\|}\sim A H^{2}\sim a_* B(H/r_{\rm g})^{2},
\label{Ep1}
\end{equation}
where we used the heuristic estimate for the GJ density $\rho_{\rm GJ}\sim a_* B/r_{\rm g}$, cf.\ Equation (\ref{nGJ}). It has the correct asymptotic values for both a non-spinning BH, $\rho_{\rm GJ}=0$ for $a_*=0$, and a rapidly spinning BH, $\rho_{\rm GJ}\sim B/r_{\rm g}$ for $a_*\sim1$ (i.e., $a\sim GM/c^{2}$). Formally, we can write
\begin{equation}
E_{\|}\simeq \beta_{\rm F} B,
\label{Ep2}
\end{equation}
where $\beta_{\rm F}\sim a_* (H/r_{\rm g})^{2}$ is a dimensionless factor. The field arises from charge separation (starvation) in a small region of size $H$ near the null surface ($\rho_{\rm GJ}=0$) due to the BZ current induced by the BH's spin in a magnetic field.

\subsubsection{Lepton acceleration}

The number of soft photons is usually large enough to produce a strong drag force, so a charge accelerated within the gap quickly attains a terminal relativistic velocity \citep{Hirotani98}. This is obtained from Equation \eqref{2} by setting ${d\Gamma}/dx=0$ and $\Gamma\gg1$. The terminal Lorentz factor is
\begin{equation}
\Gamma^2\simeq \frac{eE_\|}{\sigma_{\rm T}U_{\rm b}}.
\label{g2a}
\end{equation}
This expression is based on two assumptions, namely that (i) the Compton drag length, $\ell_{\rm C}$, is smaller than the pair-production length, $\ell_\pm$, and (ii) both are smaller than the gap size, $H$, that is, $\ell_{\rm C}\ll\ell_\pm\ll H$. For the typical conditions used in our paper, these are $\ell_{\rm C}\sim10^8$~cm, $\ell_\pm\sim10^{10}$~cm, and $H\sim10^{12}$~cm \citep{Hirotani98} for a supermassive BH of mass $M\sim 10^8 M_\odot$, with $H$ about 7\% of $r_{\rm g}$. 

In what follows, it is convenient to introduce the energy density of background photons corresponding to the Eddington luminosity:
\begin{equation}
U_{\rm Edd}=\frac{L_{\rm Edd}/c}{4 \pi r_{\rm d}^{2}}=\frac{4\pi G M m_{\rm p}}{\sigma_{\rm T}(4\pi r_{\rm d}^{2})},
\end{equation}
where $m_{\rm p}$ is the proton mass and $L_{\rm Edd}\simeq 1.25\times 10^{46} (M/10^{8}M_{\odot})$~erg/s is the Eddington luminosity. It is assumed here that the Eddington luminosity of the BH+disk system is defined at approximately the accretion disk's half-power radius, which, in turn, depends on the BH spin \citep{Page74, Fabian14} as 
\begin{equation}
r_{\rm d}\simeq [5+28(1-a_*)]r_{\rm g}\equiv f_{\rm d} r_{\rm g}.
\end{equation}
The ``disk factor,'' $f_{\rm d}$, ranges from $f_{\rm d}\sim 5$ for rapidly spinning BHs to $f_{\rm d}\sim 33$ for slowly spinning BHs. Furthermore, the region above the disk is optically thin, so photons emitted from the disk propagate isotropically. Hence, only a small fraction of these photons is directed toward the gap region and contributes to the soft photon background. Thus, the energy density of the soft disk photons entering the gap is reduced by the factor $\Omega_{\rm d}/4\pi$, where $\Omega_{\rm d}$ is the solid angle of the disk as seen from the gap location. 
Now, we normalize the energy density of the soft photons to this Eddington value:
\begin{equation}
U_{\rm b}=\epsilon_{\rm b} U_{\rm Edd}\simeq \frac{\epsilon_{\rm b}}{f_{\rm d}^2}\frac{m_{\rm p}c^{2}}{\sigma_{\rm T}r_{\rm g}}\sim(3\times10^{4}\textrm{ erg cm}^{-3}) \epsilon_{\rm b,-2}f_{\rm d,1}^{-2}M_{8}^{-1},
\end{equation}
where $M_{8}=M/(10^{8} M_{\odot})$, $f_{\rm d,1}=f_{\rm d}/10^{1}$, $\epsilon_{\rm b,-2}=\epsilon_{\rm b}/10^{-2}$, and $\epsilon_{\rm b}=(\Omega_{\rm d}/4\pi)(L_{\rm b}/L_{\rm Edd})$ is the luminosity of the background radiation field in the gap region, normalized to the Eddington luminosity. It is important to note that the factor $\epsilon_{\rm b}$ is not just the standard ``Eddington ratio" but also incorporates the disk solid angle factor $\Omega_{\rm d}/4\pi$.  

Finally, substituting $E_{\|}$ and $U_{\rm b}$ into Equation \eqref{g2a}, we obtain
\begin{equation}
\Gamma^2\simeq (eB/m_{\rm p}c)(r_{\rm g}/c)\beta_{\rm F}\epsilon_{\rm b}^{-1}f_{\rm d}^2
\sim 5\times 10^{15}\beta_{\rm F}\epsilon_{\rm b,-2}^{-1}f_{\rm d,1}^{2}M_{8}B_{5},
\label{g2}
\end{equation}
where $B_{5}=B/(10^{5}\textrm{ G})$, which is typical of accreting supermassive BH systems (e.g., as evaluated from the magnetically arrested accretion), see Section \ref{3c120}. 

\subsubsection{Cascade threshold 1: energy }

The formation of an avalanche is a threshold process driven by two independent ``reactions''. First, inverse Compton scattering produces a $\gamma$-ray photon, $e+\gamma_{\rm b}\to e+\gamma_{\gamma}$. Second, this $\gamma$-ray photon interacts with a background photon to produce an electron-positron pair, $\gamma_{\gamma}+\gamma_{\rm b}\to e^{+}+e^{-}$ (interactions of two $\gamma$-ray photons also occur, but they are rare and hence omitted). The latter is possible only if the $\gamma$-ray photon has sufficient energy for pair production. In the center-of-momentum frame, this energy condition is written as
\begin{equation}
\sqrt{h\nu_{\gamma}\,h\nu_{\rm b}}\ge m_{\rm e}c^{2}.
\end{equation}
Here, we assumed a head-on collision, which yields the minimum energy requirement. In turn, inverse Compton scattering increases the soft-photon energy by a factor of $\Gamma^{2}$, i.e., $h\nu_{\gamma}\sim\Gamma^{2}h\nu_{\rm b}$. 

Thus, the `energy threshold' constraint on the soft-photon energy reads:
\begin{equation}
h\nu_{\rm b}\ge m_{\rm e}c^{2}/\Gamma.
\label{th1}
\end{equation}

\subsubsection{Cascade threshold 2: multiplicity}

The above processes should be efficient enough that the number of produced secondaries exceeds the number of initial primary leptons. In other words, the number of produced leptons per primary lepton should exceed unity, i.e., the multiplicity should be greater than one. Otherwise, an exponential increase in the number of charges in the gap does not occur, so no avalanche can form. 

The production rates (per unit length) of the gamma photons and the leptons are
\begin{align}
d N_{\gamma}/dx & \sim \sigma_{\rm T}N_{\rm b}N_{\pm},\label{ng}\\
d N_{\pm}/dx & \sim 2\sigma_{\pm}N_{\rm b}N_{\gamma}.\label{npm}
\end{align}
where $\sigma_{\pm}$ is the pair-production cross-section and $N_{b},\ N_{\gamma},\ N_{\pm}$ are the particle densities of soft photons, hard photons, and leptons, respectively. These can be combined to yield the lepton-production equation:
\begin{equation}
d^{2} N_{\pm}/dx^{2}  \sim 2\sigma_{\rm T}\sigma_{\pm}N_{b}^{2}N_{\pm}.
\end{equation}

The multiplicity of pair production across the entire gap should exceed unity:
\begin{equation}
2\sigma_{\rm T}\sigma_{\pm}N_{\rm b}^{2}H^{2}>1.
\end{equation}
This is the `multiplicity threshold'. 

We note that Equations \eqref{ng} and \eqref{npm} are simplified versions of Equations \eqref{4} and \eqref{3}, respectively, with the functions $\eta_{\rm C}$ and $\eta_{\rm p}$ involving the cross-sections $\sigma_{\rm T},\, \sigma_{\pm}$. It is worth emphasizing that the $\gamma$-ray photon fluxes are essentially counter-propagating. Indeed, high-energy $\gamma$-ray photons are produced via Compton scattering by highly relativistic, counter-propagating electrons and positrons accelerated by the same electric field along the background $B$-field. 

For our order-of-magnitude estimates, we assume that background photons are represented by a broad-band spectrum of sufficient extent, so that Compton scattering is in the Thomson regime (i.e., neglecting Klein-Nishina corrections). This assumption aligns with our previous numerical studies of the problem and is natural for accreting systems. We should note that the case with a narrow-band (e.g., monoenergetic) spectrum should be treated more carefully. This is because pair production by IC-produced gamma-rays on a monoenergetic soft background photon population is generally in the Klein-Nishina regime, as follows from Eq. \eqref{th1}. This regime has been carefully studied by \citet{Mehlhaff+21}. To simplify our future estimates, we assume that the soft background photons can still be crudely approximated by a single energy. Although not entirely self-consistent, this assumption allows us to obtain the upper limit on the pair-production rate we are seeking and, simultaneously, remain consistent with our previous exact numerical modeling of the system \citep{Ford18,Sitarz24}.

Next, we assume that the pair-production cross-section, averaged over an isotropic distribution of seed photons, is approximately (\citealt{GS67}; see \citealt{Brown73} for corrections)
\begin{equation}
\langle \sigma_{\pm}\rangle \approx\frac{3}{4}\frac{\sigma_{\rm T}}{\Gamma^{2}_{\gamma {\rm b}}}\,\left(\ln 4\Gamma_{\gamma {\rm b}}^{2}-2\right)\simeq\frac{\sigma_{\rm T}}{\Gamma^{2}_{\gamma {\rm b}}}\,\ln\Gamma_{\gamma {\rm b}},
\end{equation}
where $\Gamma_{\gamma {\rm b}}\equiv\sqrt{h\nu_{\gamma}\,h\nu_{\rm b}}/m_{\rm e}c^{2} \simeq\Gamma\,h\nu_{\rm b}/m_{\rm e}c^{2}$ and $\Gamma_{\gamma {\rm b}}\gg 1$ are assumed. For a mono-energetic background, $N_{\rm b}\sim U_{\rm b}/h\nu_{\rm b}$. The `multiplicity threshold' now reads:
\begin{equation}
(h\nu_{\rm b})^{2}\le\sqrt{2\ln\Gamma_{\gamma {\rm b}}}(\sigma_{\rm T}/\Gamma)(m_{\rm e}c^{2})U_{\rm b}H.
\label{th2}
\end{equation}
In practical cases, the factor $\sqrt{2\ln\Gamma_{\gamma {\rm b}}}$ is on the order of a few. It is omitted from our subsequent estimates for simplicity. 

\subsection{Results}

We now combine the two avalanche-threshold conditions given by Equations \eqref{th1}, \eqref{th2}:
\begin{equation}
\frac{1}{\Gamma}\le\frac{h\nu_{\rm b}}{m_{\rm e}c^{2}}
\le\left(\frac{\epsilon_{\rm b}}{\Gamma f_{\rm d}^2}\frac{m_{\rm p}}{m_{\rm e}}\frac{H}{r_{\rm g}}\right)^{1/2}.
\end{equation}
Now, using Equation \eqref{g2} and recalling that $\beta_{\rm F}\sim(H/r_{\rm g})^{2}$, we obtain:
\begin{equation}
2\times10^{-6}\left(\frac{\epsilon_{{\rm b},-2}}{a_* f_{{\rm d},1}^{2}M_{8}B_{5} h^{2}_{{\rm g},-2}}\right)^{1/2}
\lesssim\frac{h\nu_{\rm b}}{m_{\rm e}c^{2}}
\lesssim6\times 10^{-5}\left(\frac{\epsilon_{{\rm b},-2}^3}{a_* f_{{\rm d},1}^{6}M_{8}B_{5}}\right)^{1/4}
\label{main}
\end{equation}
where $h_{\rm g}=H/r_{\rm g}$ is the dimensionless gap size, $h_{{\rm g},-2}=h_{\rm g}/10^{-2}$, and we note that the normalized disk factor is BH-spin-dependent, $f_{{\rm d},1}\simeq 0.5+2.8(1-a_*)$. 

This is an interesting and useful result. Indeed, Equation \eqref{main} shows that only soft photons within a relatively narrow optical-to-UV range can trigger a leptonic avalanche. In this case, we can expect a strong AGN jet. If the conditions are not right, e.g., the magnetic field is too low or the background spectral distribution is not optimal ($\epsilon_{\rm b}$ is too low), the cascade cannot form. For example, fields below $B \sim 1$~G should completely suppress the cascade. In this case, no plasma is produced; hence, no BZ-powered jet is present. One can argue that this mechanism can explain the observed radio-loud/radio-quiet galaxy dichotomy. 

The above calculations provide useful estimates but cannot replace actual numerical modeling. The numerical results and the instructive scalings of the system parameters are reported elsewhere \citep{Ford18, Sitarz24}.

Furthermore, a recent study of 3C 120 \citep{Zdziarski22c} quantitatively confirms and extends previous work, suggesting that an alternative mechanism is responsible for jet production in luminous sources. This is discussed in the next section. 

\section{An example of 3C 120}
\label{3c120}

3C 120 is a nearby radio galaxy at a redshift of 0.033. Its mass is $M\approx (7\pm 2)\times 10^7 M_\odot$ \citep{Grier17}, and its accretion luminosity was estimated as $L_{\rm accr}\approx 1\times 10^{45}$\,erg\,s$^{-1}$ \citep{Janiak16}, corresponding to an Eddington ratio of $\sim$0.1. It features a powerful jet \citep{Marscher02_3C120}, observed at radio, infrared, and high-energy $\gamma$-ray wavelengths. However, its X-ray emission is dominated by a hot accretion flow, as evidenced by a spectrum that includes both a power-law component characteristic of Comptonization and strong X-ray Compton reflection features \citep{ZG01, Lohfink13}. This accretion flow emission was also detected in soft $\gamma$-rays \citep{Wozniak98}. This rare combination of observational features allows the pair-production rate at the base of the jet \citep{Beloborodov99, Levinson11} to be estimated from the spectrum of photons emitted by the hot accretion flow. At the same time, the synchrotron spectrum emitted by its core jet \citep{Giommi12} allows the flow rate of the relativistic leptons responsible for that emission to be estimated. 

These quantities were estimated by \citet{Zdziarski22c} as $2\dot N_+\approx 3.9^{+6.3}_{-2.8}\times 10^{49} {\rm s}^{-1}$ and $\dot N_{\rm e}\approx 2.7_{-2.3}^{+12}\times 10^{49}\,{\rm s}^{-1}$, respectively. Within the uncertainties, they are consistent with being the same, indicating that the rate of pair production by photons from the hot accretion is fully capable of accounting for the flow of synchrotron-emitting leptons far out in the jet. Furthermore, the theoretically achievable jet power (\citealt{Davis20} and references therein) corresponds to a BZ jet launched from a magnetically arrested flow and is given by $P_{\rm j}\lesssim \dot a_*^2 M_{\rm accr}c^2= a_*^2 L_{\rm accr}/\epsilon$, where $\epsilon\sim 0.1$ is the accretion efficiency. This limits the jet power to $P_{\rm j}\lesssim 1 a_*^2 \times 10^{46}$\,erg\,s$^{-1}$. If the synchrotron-emitting leptons were mostly electrons, the jet power in the associated ions would be \citep{Zdziarski22c} $P_{\rm j}\approx 3.1^{+19.6}_{-2.8}\times 10^{47} (\epsilon/ 0.1)^{-1}{\rm erg\,s}^{-1}$, violating the theoretical constraint on the jet power by at least a factor of 10. This, in turn, implies that abundant $e^\pm$ pairs are present in the jet of 3C 120 and dominate (by number, but not by weight) its composition. This also agrees with the general estimates of \citet{Sikora20}.

We can also estimate the strength of the poloidal magnetic field threading the BH on one hemisphere. The power of a BZ jet can be written as $P_{\rm j}\approx (\phi/50)^2 a_*^2\dot M_{\rm accr} c^2$, where $\phi\lesssim 50$ (e.g., \citealt{Davis20}) is a dimensionless magnetic flux defined by $\phi\equiv 2\pi B r_{\rm g}/(\dot M_{\rm accr} c)^{1/2}$. Given that $P_{\rm j}\lesssim a_*^2 L_{\rm accr}/\epsilon$ (see above), we obtain $B\lesssim 4\times 10^5$ G, with the maximum corresponding to magnetically arrested accretion.

\subsection{Comparison with the avalanche model} 

The observational data indicate that the avalanche conditions in Equation \eqref{main} are met in 3C 120. However, is the theoretical pair-production rate in magnetospheric cascades sufficient to explain the data?

The minimum estimate is based on the GJ density. Indeed, $n_{\rm GJ}$ is the minimum plasma density required to short-circuit the parallel electric field in the rotating magnetosphere (except in a small gap region). This is also the boundary condition that $E_{\|}=0$ at $|x|=\pm H/2$ in the numerical model. In fact, this condition self-consistently determines the gap size, $H$. On the other hand, this plasma is relativistic and evacuates the magnetosphere at $v\sim c$. The corresponding particle flux density is simply $n_{\rm GJ}c$. 

The GJ density is derived from Poisson's equation, Equation \eqref{GJ}, with $E_{\rm p}\sim a_* B$. Thus, 
\begin{equation}
n_{\rm GJ}\sim a_* B/(4 \pi e r_{\rm g})\sim(1 \textrm{ cm}^{-3})a_* M_{8}^{-1}B_{5}.
\end{equation}
This is consistent with the numerical studies of \citet{Ford18, Sitarz24}. The corresponding GJ-based {\em lower limit} on the lepton production rate, equal to the total particle flux, is
\begin{equation}
\dot N_{\rm min}\sim n_{\rm GJ}c(4\pi r_{\rm g}^{2})\sim(9\times10^{37} {\textrm s}^{-1})a_* M_{8}B_{5}.
\end{equation}
For 3C 120, with $M\sim 7 \times 10^{7}M_{\odot}$ and $B\sim 4\times10^{5}$ G, we obtain $\dot N_{\rm min}\sim 3\times10^{38} \textrm{ s}^{-1}$ at $a_*\approx 1$. This is many orders of magnitude lower than the observed lepton flow rate through the jet, $\dot N_{\rm e}\sim 10^{49} \textrm{ s}^{-1}$.

We note that the actual pair-production rate may be substantially higher because many $\gamma$-ray photons produced in the gap region propagate farther and continue producing pairs far from the gap \citep{Levinson11}. The {\em upper limit} on pair production is obtained by assuming that all the energy carried by $\gamma$-ray photons is ultimately converted into $e^{\pm}$ pairs. To do this, we use the scaling relation for the total $\gamma$-ray luminosity of the cascades, derived from the self-consistent numerical solution (Equation 40 in \citet{Ford18}), which we quote here:
\begin{equation}
\frac{dL_{\gamma}}{d\Omega}\simeq(2.5\times10^{35}\ {\rm erg\,s^{-1}\,sr^{-1}})
\, a_*^{0.96}\left(\frac{M}{10^{7}\,M_{\odot}}\right)^{-2.5}
\left(\frac{B}{10^{4} \textrm{ G}}\right)^{0.95}
\left(\frac{U_{\rm b}}{10^{6}\textrm{ erg/cm}^{3}}\right)^{1.17}.
\end{equation}

The maximum pair production rate then becomes
 \begin{equation}
\dot N_{\rm max}\sim(3\times 10^{40}\textrm{ s}^{-1})\,a_* M_{8}^{-2.5} B_{5}\epsilon_{\rm b}^{1.2}f_{\rm d,1}^{-2.4},
\end{equation}
where we rounded up the exponents. For 3C 120, this yields $\dot N_{\rm max}\sim4\times10^{41}\textrm{ s}^{-1}$ even for a maximally spinning BH ($a_*=1$) with $f_{\rm d}=5$, accreting at the Eddington limit and having a huge disk solid angle $\Omega_{\rm d}=4\pi$ so that $\epsilon_{\rm b}=1$. This rate remains much too low to account for the observations.

\section{Conclusions}
\label{s:concl}

We have explored the cascade mechanism for producing e$^\pm$ pairs in an AGN environment. Such cascades are necessary for jet formation via the BZ mechanism. We have obtained crucial constraints on the magnetic field strength and the seed-photon field required for the leptonic cascade. Whether these constraints are satisfied may explain the radio-loud/radio-quiet dichotomy. 

On the other hand, we find that the production rate of e$^\pm$ pairs in the cascade mechanism is very limited. Such pairs alone may account for the particle flow rate through the jet only for very weak sources, possibly M87 or Sgr A$^*$. In luminous sources, such as bright radio galaxies or blazars, the inferred lepton flow rate from the jet synchrotron emission can be orders of magnitude higher than the maximum achievable in magnetospheric cascades. We detail this issue for the bright radio galaxy 3C 120. In that case, the cascade rate is limited to less than $4\times 10^{41}$ particles per second. This rate is several orders of magnitude lower than the value deduced from the synchrotron emission of its core jet, about $10^{49}$ particles per second. 

Thus, another mechanism for plasma production must be at work in luminous jet sources. Such a process is most likely direct pair creation in the hot accretion disk's photon field, as suggested in several previous works. This has been quantitatively confirmed in the case of 3C 120, where the synchrotron-inferred lepton flow rate was found to be compatible (within the model and observational uncertainties) with the pair-production rate inferred from its X-ray and soft $\gamma$-ray spectrum.

\section*{Acknowledgements}
MVM acknowledges support from the National Science Foundation via grant NSF PHY-2409249 and partial support via grant NSF PHY-2309135 to the Kavli Institute for Theoretical Physics (KITP). AAZ acknowledges support from the Polish National Science Center grant 2023/48/Q/ST9/00138.

\bibliographystyle{aasjournal}
\bibliography{allbib} 

\end{document}